\documentclass[a4paper,12pt]{article}
\usepackage{amsmath}
\usepackage{color}
\usepackage{epsfig}
\usepackage{subfig}
\usepackage{sidecap}
\usepackage{graphicx}
\sidecaptionvpos{figure}{c}
\begin{document}
\author{A.~S.~Khrykin}
\title{Evidence of the existence of the six-quark component of the deuteron in the energy spectra of photons emitted in proton-deuteron collisions.}
\begin{abstract}
We present evidence of the existence of the six-quark component of the deuteron ($d_{6q}$) found in the experimental photon energy spectra of the proton deuteron bremsstrahlung measured
at the proton incident energy of 200 MeV
by the Grenoble group and of 195 MeV by the Michigan state group.
A comparison of these spectra with the theoretically predicted ones revealed that the experimental spectra significantly exceed the theoretical ones.
We show that the excess of photons in the experimental spectra was caused by contributions
of new mechanisms for the photon production in $pd$ collisions, which are due to the existence of $d_{6q}$. The mechanisms comprise the process $p+d_{6q}\to p^\prime+d_{6q}^\star$, where $p^\prime$ is the scattered proton and $d_{6q}^\star$ is the $d_{6q}$ excited state, which has the total energy below the threshold for its decay by pion emission and the quantum numbers forbidden for the $NN$ system, that experiences then either the radiative decay $d_{6q}^\star\to p+n+\gamma$ or $d_{6q}^\star\to d+\gamma$.
By assuming that $d_{6q}^\star$ is the $NN$-decoupled dibaryon $d_1^\star (1956)$, we calculated the expected contribution of these mechanisms to the most precisely measured energy spectrum of photons emitted at  $90^0$ for an incident proton energy of 195 MeV.
After a simple subtraction of this contribution from the experimental spectrum, we found that the experimetally and  theoretically predicted spectra are in good agreement.
\end{abstract}
\maketitle
\section{Introduction}
In theoretical nuclear physics where the fundamental constituents of hadron matter are nucleons and mesons, the deuteron is treated as a weakly bound proton-neutron system that has the isospin $I=0$ and the spin-parity $J^P=1^+$.
This structure of the deuteron allows existence of the following mechanisms of photon production in proton-deuteron collisions at incident proton energies below the pion production threshold in free nucleon-nucleon collisions
($\pi NN$): the quasi-free radiative capture of a proton by a neutron $p+n\to d +\gamma$, the radiative capture of a proton by a deuteron $p+d\to {^3}He+\gamma$,
the reaction $p+d \to {^3}He+\pi^0 \to {^3}He+ \gamma\gamma$ and
the $pd$ bremsstrahlung that includes the quasi-free proton-proton ($pp$) and proton-neutron ($pn$) bremsstrahlung where the latter is dominant\cite{HSNak,Nak1989}.

About 30 years ago, the Grenoble group reported experimental energy spectra of hard photons ($E_\gamma > 20$ MeV) of the $pd$ bremsstrahlung emitted at several angles at the incident proton energy of 200 MeV\cite{Pinston90}.
These spectra were obtained by subtracting the expected contributions of the
$p+d\to {^3}He+\gamma$ and $p+d \to {^3}He+\pi^0$ reactions from the measured spectra
of the photons emitted from the $pd$ collisions.

Two years later, analogous energy spectra of hard photons of the $pd$ bremsstrahlung emitted at several angles at  incident proton energies of 145 MeV and 195 MeV were reported by the Michigan State group\cite{Clayton92}.
These spectra were obtained by subtracting only the expected contribution of the $p+d\to {^3}He+\gamma$ reaction from the measured spectra of the photons from the $pd$ collisions.
The reaction $p+d \to {^3}He+\pi^0$ did not make a contribution to the experimental photon spectra at an incident proton energy of 195 MeV since its threshold is 197.5 MeV. The experimental energy spectra of photons of the $pd$ bremsstrahlung for the photon emission angles of $60^0$, $90^0$ and $120^0$ were reported by both the Grenoble group for the incident proton energy of 200 MeV\cite{Pinston90} and the Michigan state group for the incident proton energy of 195 MeV\cite{Clayton92}.
The agreement between the spectra measured by these two groups is quite good for the photon emission angle of $90^0$ and satisfactory for $120^0$. For the photon emission angle of $60^0$ there is some disagreement between the spectra.

A comparison of the spectra reported by these two groups with the corresponding theoretical spectra of the $pd$ bremsstrahlung\cite{Nak1992}, which were calculated using modern meson-exchange potential models for describing the amplitude of the elementary nucleon-nucleon bremsstrahlung, has revealed that the former exceed the latter by a factor of $\sim$1.7.

As an example, Fig.1 shows a comparison between the calculated cross section of the proton-deuteron bremsstrahlung as a function of the photon energy in the laboratory frame at the incident proton energy of 195 MeV and photon emission angle of $90^0$\cite{Nak1992} and the cross section measured by the Michigan State group\cite{Clayton92}.
\begin{figure}[htb]
\begin{minipage}[c]{75mm}
\includegraphics[width=65 mm]{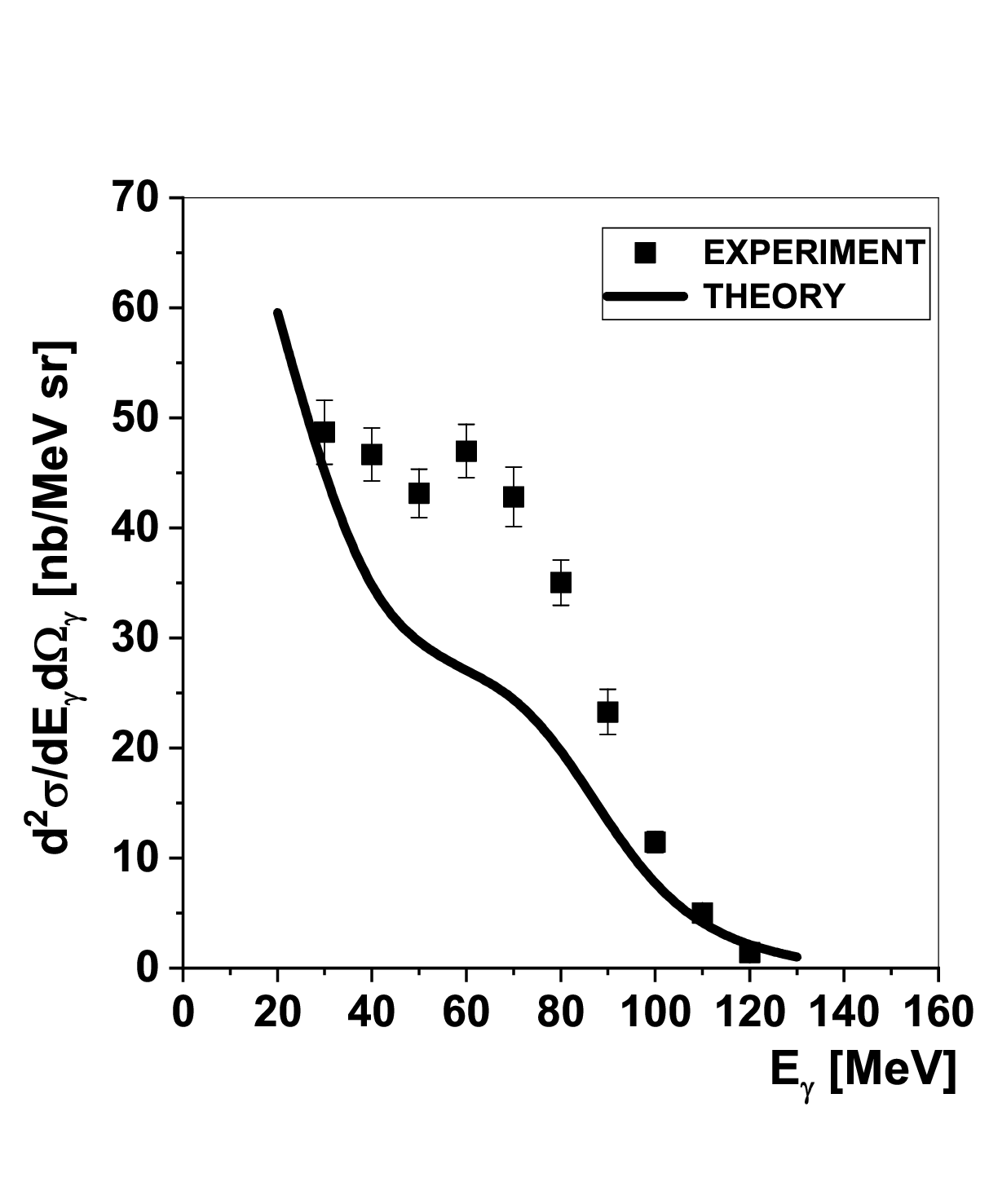}%
\end{minipage}
\hspace{\fill}
\begin{minipage}[c]{60mm}
\caption{\small{The calculated \cite{Nak1992} (solid line) and measured \cite{Clayton92} (squares) proton-deuteron bremsstrahlung cross section in the laboratory frame as a function of the photon energy $E_\gamma$ at the proton incident
energy of 195 MeV and a photon emission angle of $90^0$.}}%
\end{minipage}
\end{figure}%
The basic reason for such significant discrepancy between the experimental and theoretical photon energy spectra of the $pd$ bremsstrahlung, which are consistent with the results of the calculations for the $pn$ bremsstrahlung \cite{HSNak,Nak1989}, has not been established yet.

The fact that the experimental photon energy spectra of the $pd$ bremsstrahlung \cite{Pinston90,Clayton92}
significantly exceed the calculated ones may suggest existence of new mechanisms of the photon production in $pd$ collisions, the contributions of which were not taken into account in the calculations \cite{Nak1992}.

In this paper we propose new mechanisms of the photon production in $pd$ collisions that provide a solution to this problem. These mechanisms are due to the existence of the so-called six-quark component of the deuteron.
\section{New mechanisms of the photon production in $pd$ collisions}
In the modern theory of the strong interaction, quantum chromodynamics (QCD), the fundamental constituents of hadron and nuclear matter are quarks and gluons, and the deuteron is a color singlet six-quark system.
The structure of this dibaryon system in QCD was investigated for the first time in \cite{MatSorb1,MatSorb2},
where it was shown that the wave function of the deuteron comprises the proton-neutron ($pn$) component, the Delta-Delta ($\Delta \Delta$) component and the component in which all six quarks are in the same spatial wave function called the six-quark ($d_{6q}$) component of the deuteron. This component in the deuteron was estimated to be  $\eta\sim$ 5.6\%.

It is natural to expect that the six-quark system $d_{6q}$ has excited states $d_{6q}^\star$ which can be produced in inelastic proton-deuteron collisions
\begin{equation}
 p+d_{6q} = p' +d_{6q}^\star,
\end{equation}
where $p$ is the incident proton and $p'$ is the scattered proton.
The probability of their production in these collisions would be directly proportional to the value of $\eta$. 
One can expect that
among these states may be those $|I,J^P>$ with the isospin $I$ and the spin-parity $J^P$
which are forbidden for both singlet and triplet states of the two-nucleon systems by the Pauli exclusion principle. These are the following states:
\begin{equation}
|I,J^P> = \Biggl\{ \begin{array}{lll}
|0,0^\pm> \mbox{, } |0,1^-> \mbox{, } |0,2^-> { ,etc }\\
|1,1^\pm> \mbox{, } |1,2^+> \mbox{, } |1,3^+>{ ,etc }\\
\end{array}
\end{equation}
Certainly, any $d_{6q}^\star$ state $|I,J^P>$ from set (2), which has the total energy below the $\pi NN$ threshold,
can only experience the radiative decays $d_{6q}^\star \to \gamma+pn$ or $d_{6q}^\star \to \gamma+d$. The presence of such states in (1) leads to the existence of new mechanisms of photon production in $pd$ collisions
\begin{equation}
p+d_{6q}\to p^\prime+ d_{6q}^\star \to p^\prime+\gamma+H_f,
\end{equation}
where $H_f=pn(d)$.
A good candidate for the $d_{6q}^\star$ is the nucleon-nucleon decoupled dibaryon $d^{\ast }_{1}(1956)$ \cite{PRC64,Hyperf} which has the mass $M_{d^{\ast }_{1}}\simeq$ 1956 MeV and $|I,J^P>=|1,1^+>$.
This resonance can only decay by a photon emission and therefore it is very narrow. Its decay width is $\Gamma_{d^{\ast }_{1}} \leq$ 8 MeV.
\subsection{Contributions of new mechanisms to the total cross section of the photon production in $pd$ collisions}
Since  $\Gamma_{d^{\ast }_{1}}/M_{d^{\ast }_{1}}$ is very small, we can calculate the total cross section of reaction (3) using the narrow width approximation ($NWA$)\cite{HPilkuhn} generalized to the processes in which a decaying particle has the spin equal to one\cite{Uhlemann,Nikolas_K}.

In this approximation the total cross section for process (3) is
\begin{equation}
\sigma(pd\to p^\prime\gamma H_f)=\sigma(pd\to p^\prime d^{\ast }_{1})\times \frac{\Gamma_{{d^{\ast}_{1} \to \gamma H_f}}}{\Gamma_{d^{\ast}_{1}}},
\end{equation}
where $\sigma(pd\to p^\prime d^{\ast}_{1})$ is the total cross section of the $d^{\ast }_{1}$ resonance production,
\begin{equation}
\Gamma_{{d^{\ast}_{1} \to \gamma H_f}}=\frac{({2\pi})^4}{2E_{d^{\ast}_{1}}}\int|{\cal M}({d^{\ast}_{1}
\to \gamma H_f})|^2 d\Phi(d^{\ast}_{1}\to \gamma H_f)
\end{equation}
is the $d^{\ast}_{1}$ decay width into the final state $H_f$, where $E_{d^{\ast}_{1}}$ is the total $d^{\ast}_{1}$ energy,
${\cal M}({d^{\ast}_{1}\to \gamma H_f})$ is the matrix element for the process $d^{\ast}_{1}\to \gamma H_f$ and
$d\Phi(d^{\ast}_{1}\to \gamma H_f)$ is the element of the phase space for this process. Specifically,
\begin{equation}
d\Phi(d^{\ast}_{1} \to \gamma d)=\delta^{4}(P-k-p_d)\frac{d^3k}{(2\pi)^{3}2E_\gamma}\frac{d^3p_d}{(2\pi)^{3}2E_d}
\end{equation}
is the element of the phase space for the $d^{\ast}_{1}\to \gamma d$ decay and
\begin{equation}
d\Phi(d^{\ast}_{1} \to \gamma pn)=\delta^{4}(P-k-p_p-p_n)\frac{d^3k}{(2\pi)^{3}2E_\gamma}\frac{d^3p_p}{(2\pi)^{3}2E_p}
\frac{d^3p_n}{(2\pi)^{3}2E_n}
\end{equation}
is the element of the phase space for the $d^{\ast}_{1}\to \gamma pn$ decay and $\Gamma_{d^{\ast}_{1}}=\Gamma_{{d^{\ast}_{1} \to \gamma pn}}+\Gamma_{{d^{\ast}_{1} \to \gamma d}}$ is the total $d^{\ast}_{1}$ decay width.

In \cite{Uhlemann} it was proved that
for total and sufficiently inclusive differential rates of an arbitrary resonant decay or scattering processes with an on-shell intermediate state decaying via a cubic or quartic vertex the NWA is of $\mathcal{O}(\Gamma)$.
\subsection{Cross section for the $d^{\ast }_{1}$ production in $pd$ collisions}
In \cite{Filkov2} the differential cross section $\frac{d\sigma}{d\Omega_{p^\prime}}$ for the $d^{\ast }_{1}$ dibaryon production in proton deuteron collisions $p+d \to p^\prime+ d^{\ast }_{1}$ was obtained. It is 
\begin{equation}
\frac{d\sigma}{d\Omega_{p^\prime}}=\eta\frac{\textsf{g}_1^2}{4\pi} F_1[(M_{d_1^\star}^2+m_d^2-t)^2-4m_d^2M_{d_1^\star}^2],
\end{equation}
where
$\frac{\textsf{g}_1^2}{4\pi}$ is the coupling constant in the vertex for the transition of the $d_{6q}$ state into the $d_1^\star$ state via the pion exchange mechanism $d_{6q}+\pi \to d_1^\star$, and $t=(p-p^\prime)^2$ is the four momentum transfer squared, with the four momenta of the incident $p$ and final $p^\prime$ protons, $m_d$ is the deuteron mass
\begin{equation}
F_1=-\frac{1}{4} \frac{\textsf{g}_{\pi NN}^2}{4\pi} \frac{tF}{(t-m_\pi^2)^2},
\end{equation}
where $\frac{\textsf{g}_{\pi NN}^2}{4\pi}=14.6$, $m_\pi$ is the pion mass,  and
\begin{equation}
F=\frac{8}{3}\big(\frac{|\vec{p^\prime}|}{M_{d_1^\star}}\big)^2\frac{1}{m_d|\vec{p}
|[(m_d+E)|\vec{p^\prime}|-|\vec{p}|E^\prime*\cos\theta_{p\prime}},
\end{equation}
where $E$ and $E^\prime$ are the energies of the incident and scattered protons.
The coupling constant $\frac{\textsf{g}_1^2}{4\pi}$ is unknown, we set it equal to 1.
\subsection{Matrix elements for the process $d^{\ast}_{1}\to \gamma H_f$}
The general expression for the square of the matrix element for the process $A\to \gamma B$, where $A$ and $B$ are hadrons, was obtained in \cite{Lautrup}. It is
\begin{equation}
|{\cal M}(A\to \gamma B)|^2=(p_A \cdot k)^2\cdot2M_T(k^2),
\end{equation}
where $p_A$ and $p_B$ are the four momenta of the initial and final hadrons, $k=p_A-p_B$ is the four momentum of the emitted photon, and $M_T(k^2)$ is the transverse spectral function\cite{Lautrup} which is a real function of $k^2$. The kinematical region for $k^2$ is $0 \le k^2 \le (m_A-m_B)^2$, where $m_A$ and $m_B$ are the invariant masses of the initial and the final hadrons, respectively.
Using Eq.(11), we obtain
\begin{equation}
|{\cal M}({d^{\ast}_{1}
\to \gamma d})|^2=(p_{d^{\ast}_{1}}\cdot k)^2\cdot 2M_T(k^2),
\end{equation}
where $k=p_{d^\ast_1}-p_d$, with $p_{d^\ast_1}$ and $p_d$ being the four momenta of the $d^{\ast}_{1}$ and the final deuteron.

The process $d_1^\star \to \gamma pn$ involves two subprocesses. The subprocess of the $d^{\ast}_{1}$ radiative decay into the $pn$ system and the subprocess of the final state interaction ($FSI$) of the nucleons of this system.
The square of the total matrix element for this process $|{\cal M}({d^{\ast}_{1}\to \gamma pn})_{tot}|^2$ was obtained in the spirit of the Watson-Migdal approximation prescription\cite{Watson,WatMig,ShMos,GoldWat} according to which it was factorized as
\begin{equation}
|{\cal M}({d^{\ast}_{1}\to \gamma pn})_{tot}|^2=|{\cal M}({d^\ast_1 \to \gamma pn})|^2\cdot|{\cal M}_{FSI}|^2,
\end{equation}
where $|{\cal M}({d^{\ast}_{1}\to \gamma pn})|^2$=$(p_{d^\ast_1}\cdot k)^2\cdot 2M_T(k^2)$ is the square of the matrix element for the $d^\ast_1$ radiative decay, $k=p_{d^\ast_1}-p_{pn}$ with
$p_{pn}=p_p+p_n$, where $p_p$ and $p_n$ are the four momenta of the final proton and neutron, respectively, and $|{\cal M}_{FSI}|^2$ is the final state interaction enhancement factor, which is given by the inverse Jost function $J_l^{-1}(p_r)$, where $l$ and $p_r$ are the partial wave and the relative momentum of the two final nucleons.

In view of the expected smallness of the relevant relative energies of the nucleons of the final $pn$ system of the $d_1^\star \to \gamma pn$ process, one should expect that the relative orbital angular momentum of the two nucleons of this system $l_r$ is dominantly zero. Therefore, the  $pn$ system can be in the singlet $^1S_0$ or the triplet $^3S_1$ state, and the total enhancement factor is
\begin{equation}
|{\cal M}_{FSI}|^2=1/4|{\cal M}_{FSI}^S|^2+3/4|{\cal M}_{FSI}^T|^2,
\end{equation}
where
$|{\cal M}_{FSI}^S|^2$ and $|{\cal M}_{FSI}^T|^2$ are the enhancement factors for the radiative transition $|1,1^+> \to |1,0^+>$ and $|1,1^+> \to |1,1^+>$, respectively.

The enhancement factors $|{\cal M}_{FSI}^S|^2$  and $|{\cal M}_{FSI}^T)|^2$ were calculated using the s-wave Jost function, which was obtained in \cite{ShMos}.
This function is based on the effective range expansion equation
\begin{equation}
p_r cot(\delta_0)=(1/a_0)+(1/2)r_0p_r^2,
\end{equation}
where $p_r$ is the relative momentum of two outgoing nucleons,
$\delta_0$ is the S-wave $pn$ phase shift, and $a_0$ and $r_0$ are the scattering length and effective range parameters, respectively. The corresponding Jost function is
\begin{equation}
J_0^{S(T)}(p_r)=\frac{1/a_0+(r_0/2)p_r^2-\imath p_r}{(p_r^2+\alpha^2)r_0/2},
\end{equation}
where
\begin{equation}
\alpha=(1/r_0)[1+(1+2r_0/a_0)^{1/2}].
\end{equation}
The enhancement factor $|{\cal M}_{FSI}^{S(T)}|^2$ is given by
\begin{equation}
|{\cal M}_{FSI}^{S(T)}|^2=\frac{1}{|J_0^{S(T)}(p_r)|^2}
\end{equation}
In our calculations we have used the following values for the scattering length $a_0$ and effective range $r_0$ parameters:
$a_0(\mbox{singlet})=23.715$ fm, $r_0(\mbox{singlet})=2.750$ fm, $a_0(\mbox{triplet})=-5.4112$ fm, $r_0(\mbox{triplet})=1.7436$ fm \cite{EFRAP,ShMos,NOYES}.
\subsection{Contribution of the process $pd\to p^\prime\gamma H_f$ to the spectrum of photons emitted in $pd$ collisions\cite{Pinston90,Clayton92}}
The contribution of process (3) to the differential spectrum of the photons  emitted in $pd$ collisions $\frac{d^2\sigma}{dE_\gamma d\Omega_\gamma}$\cite{Pinston90,Clayton92} can be written in the form
\begin{equation}
\frac{d^2\sigma(pd\to p^\prime\gamma H_f)}{dE_\gamma d\Omega_\gamma}\simeq \sigma(pd\to p^\prime d^{\ast }_{1})_{H_F}\frac{1}{\Gamma_{d^{\ast}_{1}}}\frac{d}{dE_\gamma}\frac{1}{\Delta\Omega_\gamma}\Gamma(d^{\ast }_{1}\to\gamma H_f)
\end{equation}
where $\sigma(pd\to p^\prime d^{\ast }_{1})_{H_F}$ is the total cross section of the $d^{\ast }_{1}$ production restricted to the events each of which has a photon from its decay with the energy $E_\gamma \ge 20$ MeV emitted into the solid angle $\Delta\Omega_\gamma$ covered by the $\gamma$-detector\cite{Clayton92}. Because in the experiments \cite{Pinston90,Clayton92} only photons with the energy $E_\gamma \ge 20$ MeV were detected, the cross section $\sigma(pd\to p^\prime d^{\ast }_{1})_{pn}$ is not equal to the cross section $\sigma(pd\to p^\prime d^{\ast }_{1})_d$.

The spectrum (19) is not directly calculable since the quantity $\Gamma(d^{\ast }_{1}\to\gamma H_f)$ contains the quantity $M_T(k^2)$ the value of which is unknown.

To obtain this spectrum, we rewrite expression (19) in the form
\begin{equation}
\frac{d^2\sigma(pd\to p^\prime\gamma H_f)}{dE_\gamma d\Omega_\gamma}\simeq \sigma(pd\to p^\prime d^{\ast }_{1})_{H_f}\frac{\Gamma(d^{\ast }_{1}\to\gamma H_f)}
{\Gamma_{d^{\ast}_{1}}}\times
\end{equation}
\begin{equation}
\frac{1}{\Gamma(d^{\ast }_{1}\to\gamma H_f)}\frac{d}{dE_\gamma}\frac{1}{\Delta\Omega_\gamma}\Gamma(d^{\ast }_{1}\to\gamma H_f)=\sigma(pd\to p^\prime\gamma H_f)\times W(E_\gamma)\nonumber,
\end{equation}
were
\begin{equation}
W_{H_f}(E_\gamma)=\frac{1}{\Gamma({d^{\ast }_{1}}\to \gamma H_f)}\frac{d}{dE_\gamma}\frac{1}{\Delta\Omega_\gamma}{\Gamma({d^{\ast }_{1}}\to \gamma H_f)}.
\end{equation}
is the normalized-to-unity energy spectrum of photons in the energy range 20 MeV $\le$ $E_\gamma \le$ 170 MeV emitted from the process $d^{\ast }_{1}\to \gamma H_f$ into the solid angle $\Delta\Omega_\gamma$ covered by the $\gamma$-detector \cite{Clayton92}. The quantity $W_{H_f}(E_\gamma)$ does not depend on $M_T(k^2)$ and is calculable.

The energy spectra $W_{H_f}(E_\gamma)$ and the values of the corresponding cross sections $\sigma(pd\to p^\prime d_1^\star)|_{H_f}$ for $H_f=pn(d)$ were calculated for the incident proton energy of 195 MeV and the photon emission angle of $90^0$.

The value of the cross section $\sigma(pd\to p^\prime d_1^\star)|_{H_f}$  is proportional to the value of the constant $C_{6q}=\eta*\frac{q_1^2}{4\pi}$. In our calculation this constant was set equal to 0.01.
The calculations were carried out by the Monte Carlo method.
It was assumed in these calculations that in the CMS system the angular distribution of the scattered proton $p^\prime$ of the reaction $p+d_{6q}=p^\prime+d^{\ast}_{1}$  is isotropic. The four momenta of the particle of the $d^{\ast}_{1}$ decay
were calculated using the GENBOD event generator\cite{Genbod}.
As a result, we obtained $\sigma(pd\to p^\prime d_1^\star)|_{pn}$ = 0.307 nb and $\sigma(pd\to p^\prime d_1^\star)|_{d}$ = 0.412 nb.

The quantities of interest are the differential cross sections of the photon production in the $pd$ collisions $\frac{d^2\sigma(pd\to p^\prime \gamma H_f)}{dE_\gamma d\Omega_\gamma}$ due to the process $(pd\to p^\prime \gamma H_f)$ for $H_f=pn$ and $H_f=d$.
In order to find the magnitudes of these quantities, each calculated spectrum $W_{H_f}(E_\gamma)$ was fitted to the experimental spectrum\cite{Clayton92} by varying only one parameter $\sigma(pd\to p^\prime \gamma H_f)$.
As a result, we found that $\sigma(pd\to p^\prime \gamma pn)\simeq 7.368$ nb and $\sigma(pd\to p^\prime \gamma d)\simeq 0.5$
nb. The results of fitting are shown in Fig.2
\begin{figure}[htb]
\begin{minipage}[c]{75mm}
\includegraphics[width=65 mm]{{{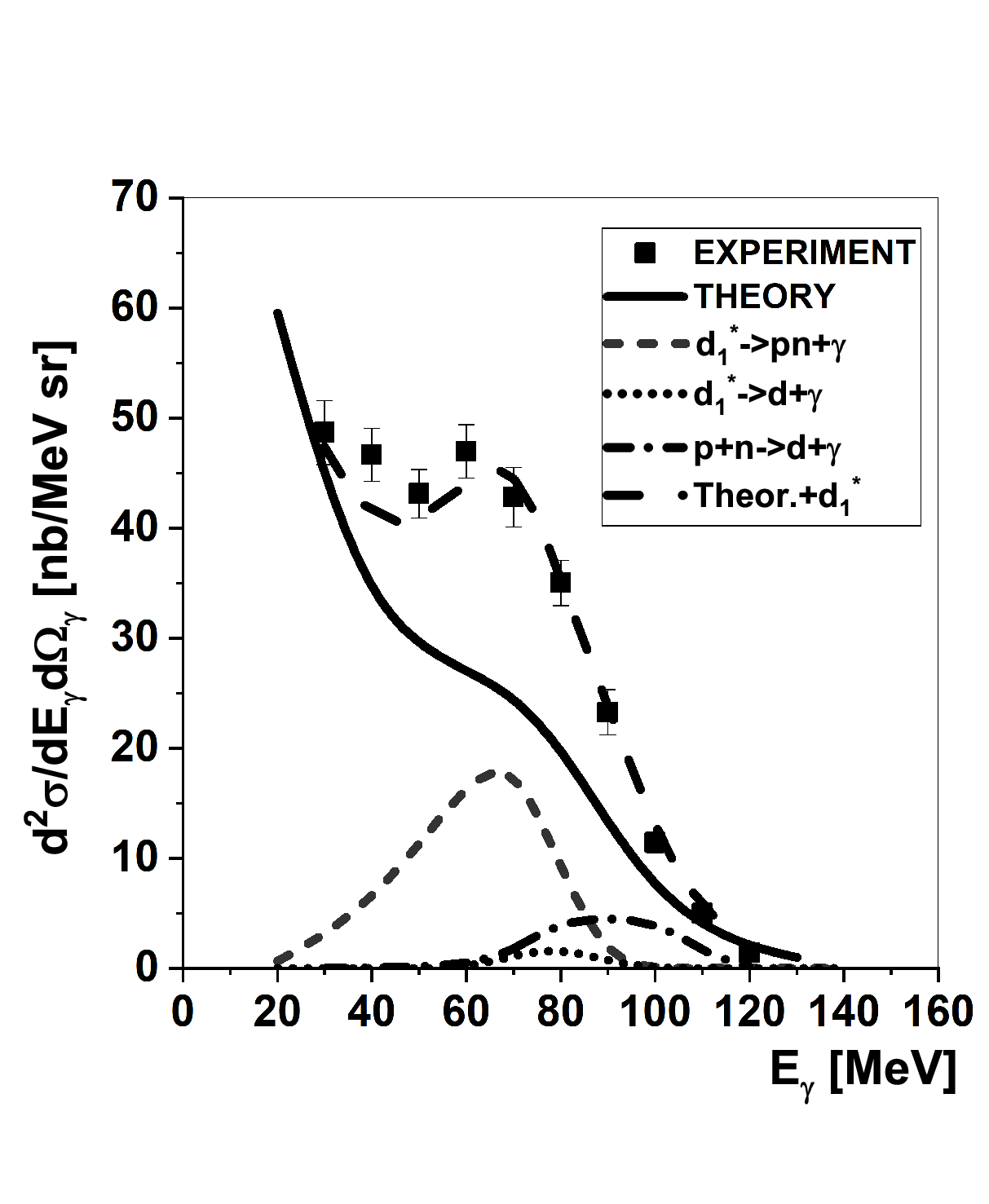}}}
\end{minipage}
\hspace{\fill}
\begin{minipage}[c]{65mm}
\caption{\small{The calculated \cite{Nak1992} (black solid line) and the measured \cite{Clayton92} (black squares) proton-deuteron bremsstrahlung cross section in the laboratory frame as a function of the photon energy $E_\gamma$ at the proton incident energy of 195 MeV and the photon emission angle of $90^0$. The short-dashed line corresponds to the contribution of the $d^\star_1 \to \gamma pn$ decay and the dotted line corresponds to the contribution of the $d^\star_1 \to \gamma d$ decay. The dash-dotted line corresponds to the contribution of the $p+n=d+\gamma$ reaction. The dashed line is the sum of these contributions}}%
\end{minipage}
\end{figure}

The obtained data enable us to estimate the value of the constant $C_{6q}=\eta*\frac{q_1^2}{4\pi}$.
Since the cross section $\sigma(pd\to p^\prime \gamma pn) = \sigma(pd\to p'd_1^\star)_{pn}\frac{\Gamma(d_1^\star\to\gamma pn)}{\Gamma_{d_1^\star}}$ is considerably larger than $\sigma(pd\to p^\prime \gamma d) = \sigma(pd\to p'd_1^\star)_{d}\frac{\Gamma(d_1^\star\to\gamma d)}{\Gamma_{d_1^\star}}$, then $\Gamma(d_1^\star\to\gamma pn) >> \Gamma(d_1^\star\to\gamma d)$ and $\Gamma_{d_1^\star}\simeq \Gamma(d_1^\star\to\gamma pn)$.
Therefore, the cross section $\sigma(pd\to p^\prime \gamma pn)\simeq \sigma(pd\to p^\prime {d_1^\star})_{pn}^r$, where the  quantity $\sigma(pd\to p^\prime {d_1^\star})_{pn}^r$ corresponds to the real value of the quantity $C_{6q}$ which we call $C_{6q}^{r}$. We can now write
\begin{equation}
\frac{C_{6q}^{r}}{C_{6q}}=\frac{\sigma(pd\to p^\prime {d_1^\star})_{pn}^{r}}{\sigma(pd\to p^\prime {d_1^\star})_{pn}^{cal}}=\frac{7.368\ nb}{0.307\ nb}=24.
\end{equation}
Since $C_{6q}=0.01$, we get $C_{6q}^{r}\simeq 0.24$. If we assume that $\eta$=0.06, then the constant $\frac{q_1^2}{4\pi}$ = 4.
\section{Conclusions}
We have shown that the considerable excess of the experimental photon energy spectra\cite{Clayton92} of the $pd$ bremsstrahlung process over the theoretical\cite{Nak1992} ones arises from contributions of new mechanisms for photon production in proton deuteron collisions which are not included in the calculations. These new mechanisms of photon production in proton deuteron collisions are induced by the existence of the six-quark component of the deuteron. The inclusion of these mechanisms in calculations permits obtaining a good agreement between the experimental and calculated  photon energy spectra of the $pd$ bremsstrahlung.
\section{APPENDIX}
In this appendix we calculate the contribution of the quasi-free $pn$ radiative capture process to the differential cross section for production of the photons with the energy $E_\gamma \ge 20 MeV$ in $pd$ collisions $d^2\sigma(pn\to \gamma d)/dE_\gamma d\Omega_\gamma$.
The quasi-free radiative $pn$ capture process is
\begin{equation}
p+n\to d+\gamma,
\end{equation}
where $p$ is the incident proton, and $n$ is the neutron in the deuteron. The differential cross section of the photon production in process (23) is given by
\begin{eqnarray}
\frac{d^2\sigma(pn\to\gamma d)}{dE_\gamma d\Omega_\gamma}&=&\frac{d}{dE_\gamma}\frac{d}{d\Omega_\gamma}\frac{1}{(2\pi)^{2} 4F}\int \overline{|{\cal M}({p_p+p_n\to\gamma d})|^2} d\Phi_2\\
d\Phi_2&=&\frac{d^3\vec{k}}{2E_\gamma}\frac{d^3\vec{p}_d}{2E_d}\delta^4(p_p+p_n-k-p_d)
\end{eqnarray}
where $F=\sqrt{(p_pp_n)^2-(m_pm_n)^2}$ is the flux factor, $p_p=(E_p,\vec{p}_p)$ and $p_n=(E_n,\vec{p}_n)$ are the four momenta of the incident proton and participant neutron, respectively, $k=(\omega,\vec{k})$ and $p_d=(E_d,\vec{p}_d)$ are the four momenta of the emitted photon and the deuteron, respectively, and $\overline{|{\cal M}({p_p+p_n\to\gamma d})|^2}$ is the square of the Lorentz invariant matrix element for process (23) averaged over the initial and summed over the final spin states of the particle of this process and polarizations of the emitted photon. In our calculation we used the matrix element (9) which for process (23) is equal to
\begin{equation}
\overline{|{\cal M}({p_p+p_n\to\gamma d})|^2}=(P_{pn}\cdot k)^2\cdot M_T(k^2)=\omega^2[E_{pn}-|\vec{P}_{pn}| \cdot cos(\theta_\gamma)]^2,
\end{equation}
where $P_{pn}=p_p+p_n$.

In the center-of-mass system the energy of photons emitted from process (21)  is
\begin{equation}
\omega(pn\to\gamma d)=(s_{pn}-m_d^2)/2\sqrt{s_{pn}},
\end{equation}
where $s_{pn}=P_{pn}^2$ is the invariant mass square of the $pn$ system.
\begin{equation}
s_{pn}=(p_p+p_n)^2=m_p^2+m_n^2+2\vec{p}_p\cdot\vec{p}_n.
\end{equation}
The Fermi momentum distribution of the nucleon ($p_N$) in the deuteron was measured\cite{Fmomment}
and was fitted with the function  $f(p_N)$ given by \cite{FmomPar}
\begin{equation}
f(p_N)=1.7716\times10^{-5}
p_N^2(e^{-0.00063p_N^2}+0.201e^{-0.026p_N}
+0.0119e^{[-(p_N-77.7)/38.8]^2}).
\end{equation}
The function $f(p_N)$ is normalized so that
\begin{equation}
\int\limits_0^\infty f(p_N)dp_N=1  \nonumber
\end{equation}
The nucleon momentum distribution inside the deuteron given by expression (27) compared with the measured momentum distribution is presented in Fig.3.
\begin{figure}[htb]
\begin{minipage}[c]{75mm}
\includegraphics[width=50 mm]{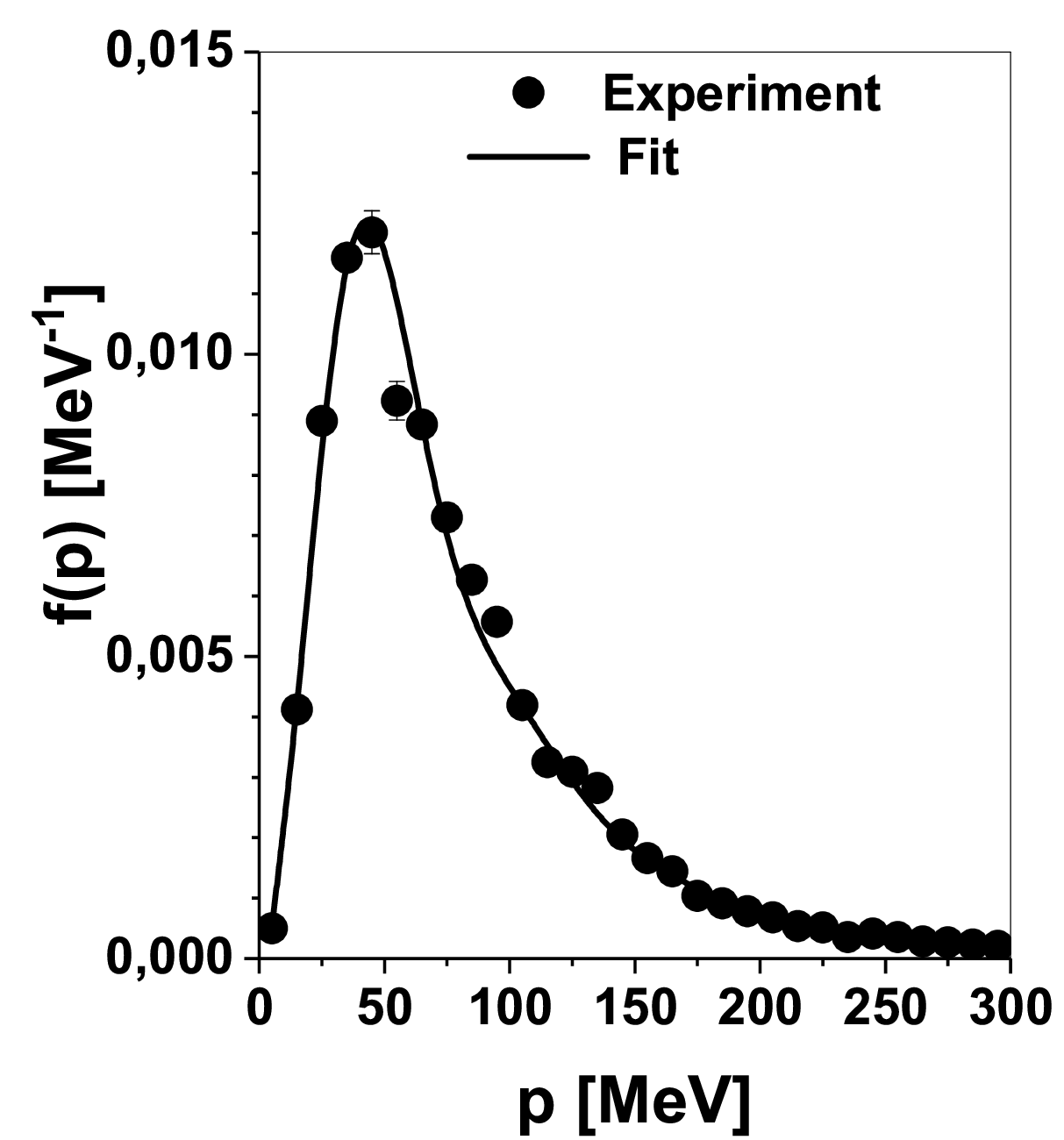}%
\end{minipage}
\begin{minipage}[c]{60mm}
\caption{The measured Fermi momentum distribution (black solid points) compared with the parametrization (black solid line)}%
\end{minipage}
\end{figure}%

Expression (24) is not calculable since it includes the unknown constant $M_T(k^2)$. Therefore, we calculated
the normalized-to-unity spectrum of photons of the process (23) $W(pn\to\gamma d)$ which is given by
\begin{equation}
W(pn\to\gamma d)=\frac{d^2\sigma(pn\to\gamma d)}{dE_\gamma d\Omega_\gamma}/\sigma(pn\to\gamma d)
\end{equation}
The calculation was carried out by the Monte Carlo method.
Then the quantity $\sigma(pn\to\gamma d)\times W(pn\to\gamma d)$ was fitted to the experimental spectrum\cite{Clayton92}
by varying only one parameter $\sigma(pn\to\gamma d)$. The result of the fit is shown in Fig.2 by the dash-dotted line.

\end{document}